\documentclass[reprint,amsmath,amssymb,aps,pra
]{revtex4-2}

\usepackage{graphicx}
\usepackage{dcolumn}
\usepackage{bm}
\usepackage[utf8]{inputenc}
\usepackage{nicefrac}
\usepackage{natbib}
\usepackage{amsmath}
\usepackage[dvipsnames]{xcolor}
\usepackage[T1]{fontenc}
\usepackage{mathptmx}
\usepackage{etoolbox}
\usepackage{booktabs}
\usepackage{hyperref}
\usepackage{soul}
\newcommand{\RNum}[1]{\uppercase\expandafter{\romannumeral #1\relax}}
\usepackage{xcolor}
\usepackage[dvipsnames]{xcolor}

\makeatletter

\def\@email#1#2{%
 \endgroup
 \patchcmd{\titleblock@produce}
  {\frontmatter@RRAPformat}
  {\frontmatter@RRAPformat{\produce@RRAP{*#1\href{mailto:#2}{#2}}}\frontmatter@RRAPformat}
  {}{}
}%
\makeatother

\begin{document}
\preprint{APS/123-QED}

\title{Continuous-wave laser source at the 148\,nm  nuclear transition of Th-229}

\author{{V. Lal}$^1$}
\author{M. V. Okhapkin$^1$} 
\author{J. Tiedau$^1$}
\author{N. Irwin$^1$}
\author{V. Petrov$^2$}
\author{E. Peik$^{1}$}
\email{email: ekkehard.peik@ptb.de}
\affiliation{$^1$ Physikalisch-Technische Bundesanstalt, 38116 Braunschweig, Germany}
\affiliation{$^2$ Max-Born-Institute for Nonlinear Optics and Ultrafast Spectroscopy, Max-Born-Str. 2A,  12489 Berlin, Germany}

\vspace{10pt}

\begin{abstract} 
A continuous-wave laser source at 148.4\,nm  based on second-harmonic generation in randomly quasi-phase matched strontium tetraborate, SrB$_4$O$_7$, is demonstrated. It provides 1.3\,$_{-0.6}^{+0.7}$\,nW of VUV power in a single pass for an incident UV laser power of 325\,mW. The laser system is developed for the resonant laser excitation of the $^{229}$Th nucleus to its low-energy isomeric state.
For a frequency-stabilized laser system we expect to reach similar VUV power spectral densities as in  previous pulsed laser excitation experiments of the nuclear transition in $^{229}$Th-doped crystals. 
\end{abstract}

\maketitle
The recently achieved laser excitation of a low-energy nuclear transition in the isotope Thorium-229 \cite{tiedau2024laser,elwell2024laser,zhang2024frequency,zhang2024229thf4} has opened a novel domain of nuclear laser spectroscopy. For the first time it has become possible to resonantly drive a radiative nuclear transition with a table-top laser system. Several possible applications have been proposed, including an optical clock with very high accuracy and stability \cite{peik2003nuclear,campbell2012single} and a nuclear $\gamma$-ray laser \cite{tkalya2011proposal}. 
Methods that have been developed in atomic and molecular laser spectroscopy can now be applied to a nuclear system that offers the advantage of an insensitivity to external perturbations predominantly affecting the electron shell. On the other hand, the fact that the nuclear excitation energy of 8.4\,eV lies in the same range as excitations of valence electrons may also be used in studies of specific coupling mechanisms between the electron shell and the nucleus, such as an electronic bridge process \cite{Tkalya1996, matinyan1998lasers}.  
With a long lifetime  $\approx~10^3$\,s and a central wavelength of 148.4\,nm, exploiting the full potential of the $^{229}$Th transition for high-resolution laser Mössbauer spectroscopy or for an optical nuclear clock poses a challenge for the development of ultra-narrow linewidth lasers \cite{Beeks2021}. In the pioneering experiments with laser sources based on four-wave mixing of ns laser pulses, the obtained linewidths have been limited to the GHz range \cite{tiedau2024laser,elwell2024laser} while spectral widths of $^{229}$Th lines in the order of 300\,kHz \cite{zhang2024frequency} and 30\,kHz \cite{ooi2025frequency} were observed with the 7$^{th}$ harmonic of a fs-laser frequency comb.

The nonlinear frequency conversion efficiency in gases requires high peak intensities which will either lead to excessive spectral linewidths or, in case of frequency combs, to rather low power in a single tooth and an ac Stark shift caused by all comb modes 
\cite{udem2007frequency}. A very promising solution for the generation of narrow-linewidth vacuum ultraviolet (VUV) light is the frequency conversion in acentric optical crystals with second order nonlinear susceptibility. However, no such crystals exist which are simultaneously transparent at 148 nm and possess sufficient birefringence for phase-matched second-harmonic generation (SHG) at this wavelength. The shortest SHG wavelength of 158.9\,nm, recently reported in NH${_4}$B${_4}$O${_6}$F (ABF), still does not permit access to the thorium isomeric state excitation energy of 8.4\,eV (148.4\,nm)  \cite{zhang2025full}. For sum-frequency generation (SFG) our calculations show that from the available crystals only BaMgF4 (BMF) is applicable for the target wavelength of 148\,nm, but this approach is limited by the very low nonlinear coefficient $<$\,0.04\,pm/V of the material \cite{villora2009birefringent}. 
 
When birefringent phase-matching is impossible, quasi-phase matching (QPM) in structured crystals is often the solution to generate the desired wavelength in a three-wave nonlinear process.
However, attempts to fabricate QPM
structures (e.g., \cite{shao2022angular,buchter2001periodically, herr2023fanout}) have at present not yielded successful generation of VUV radiation at 148\,nm.

In random QPM (RQPM) SHG, waves generated with random phases add up to a SHG power that grows linearly with the crystal length \cite{baudrier2004random}.
The shortest wavelength of 121\,nm ever reported from a second-order nonlinear optical process was achieved in a RQPM strontium tetraborate SrB$_4$O$_7$ (SBO) crystal
using a femtosecond laser amplifier  \cite{trabs2016generation}.
The structure, originated from a Czochralski growth process
\cite{aleksandrovsky2011deep}, contains spontaneously poled domains with opposite orientation normal to the \textit{a}-axis of the non-ferroelectric SBO.
Strontium tetraborate offers a high nonlinear optical coefficient d$_{33}$ (1.5 - 3.5\,pm/V) \cite{Petrov:04,trabs2016generation}, a good resistivity to optical damage, and chemical stability. 
Therefore, RQPM  SHG in SBO crystals can be used for a nonlinear optical conversion of continuous-wave (CW) radiation to the vacuum ultraviolet spectral region. 

\begin{figure}[t]
\centering
\includegraphics[width=0.45\textwidth]{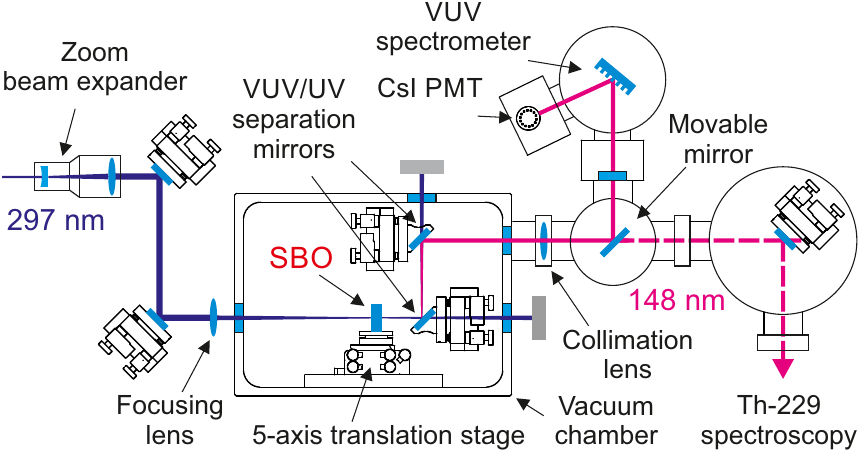}
\caption{\label{fig:spectroscopy} Experimental setup for SHG at 148.4\,nm. UV light at 296.8\,nm is focused into the RQPM SBO crystal that generates VUV light. The temperature stabilized crystal is mounted on a 5-axis translation stage. Two dichroic mirrors, a VUV grating and a solar blind CsI PMT are used for the detection of the VUV power. The VUV radiation can be directed either to the spectrometer or to the spectroscopy beam line by a mirror mounted on a vacuum compatible translation stage.}
\end{figure}

Here, we report the development of an  all-solid-state CW laser system for the $^{229}$Th nuclear transition, based on three sequential SHG steps starting from a diode laser at 1187\,nm. In particular, we demonstrate successful frequency doubling of CW laser radiation from 296.8\,nm to 148.4\,nm in a RQPM SBO crystal.

The setup for detecting VUV radiation from SHG is designed to be sensitive to very low light powers on the fW level. The SHG process is driven by a commercial frequency quadrupled CW laser (Toptica TA-FHG pro) with an output power of $\approx$\,400\,mW at a wavelength of 296.8\,nm and a specified linewidth in the order of 100\,kHz.
In this work, we are using the laser without active frequency stabilization.

For SHG we use the same uncoated SBO crystal with irregular domain structure described in Ref.~\cite{trabs2016generation}. The laser radiation is focused into the SBO sample with dimensions of 0.9\,$\times$\,6.4\,$\times$\,5.6\,mm$^3$ placed in a vacuum chamber for the final conversion to 148.4\,nm (see Fig.~\ref{fig:spectroscopy}).
The domains have a random thickness from <\,0.1 to >\,100\,$\mu$m along the light propagation direction (crystal \textit{a}-axis). The polarization of the incident laser radiation is oriented parallel to the crystal \textit{c (z)}-axis. The crystal holder provides a rotation of the SBO around the \textit{c} and \textit{b}-axes. It is mounted on a 5-axis vacuum compatible translation stage for fine optimization of the crystal position and orientation. The SBO crystal is temperature stabilized via a Peltier thermoelectric module. 
At 296.8\,nm we observe transmission losses in this sample on the order of a few percent in addition to the reflection losses of $\approx$\,8\,$\%$. 
The losses can partially be attributed to a minor surface damage from earlier experiments with this crystal under high-intensity UV irradiation. 
An orientation of the SBO crystal at the Brewster angle is not optimal for SHG due to the increase in period thickness of the complex domain structure.
Because of the high total losses for the fundamental radiation in combination with quite small areas of efficient RQPM (see below), we do not consider cavity-enhanced SHG for this specific crystal.

\begin{figure}[h]
\centering
\includegraphics[width=\linewidth]{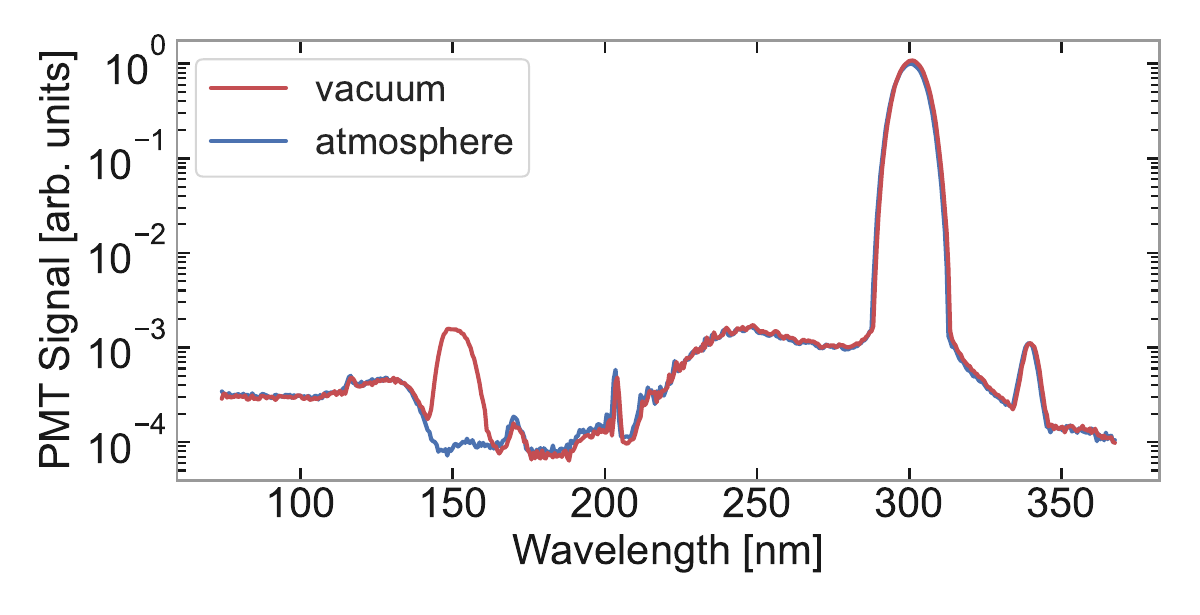}
\caption{\label{fig:crystal_scan}
The recorded PMT signal showing fundamental and second-harmonic spectra. The SHG signal at 148\,nm vanishes under atmospheric conditions due to high absorption of the VUV light. The peaks between 170 and 340\,nm are residuals of the UV radiation.}
\end{figure}

To find an optimal beam waist in the SBO crystal we vary the laser beam diameter with a zoom beam expander and focus the beam into the crystal with a 200\,mm lens. The total power losses of the optical elements used to direct the UV beam to the SBO sample are $\approx$\,20$\,\%$.
Two dichroic mirrors are used for the separation of the VUV and UV beams and to direct the VUV radiation to a detector. The SHG vacuum chamber is separated from the other parts of the vacuum system by a MgF$_2$ viewport with a transmission of $\approx$\,90$\,\%$ at 150\,nm. For the collimation of the VUV radiation we are using a positive MgF$_2$ lens.
Finally, the light is directed into a VUV-Spectrometer (HP Spectroscopy easyLIGHT) equipped with a CsI photon multiplier tube (PMT) (Hamamatsu R8487) for the detection of the VUV radiation. The spectrometer is separated from the spectroscopy beam line by a second MgF$_2$ viewport with the same transmission.

The main challenge for the SHG detection is obtaining sufficient suppression of the fundamental radiation. This requirement can be estimated from the ratio of initial UV-power ($\approx$\,0.3\,W) compared to a weak VUV signal in the fW range (under non-optimized conditions), which corresponds to a suppression factor of 14 orders of magnitude. 
In addition to a solar-blind CsI PMT \cite{philipp1956photoelectric} with a low efficiency in the UV, we use two dichroic VUV/UV separation mirrors in combination with a spectrometer grating for the extraction of the VUV signal from the UV background.
The dichroic mirrors are characterized in the UV with a calibrated photon flux resulting in an estimated attenuation factor of about 100. The combined reflection of the two mirrors in the UV is measured to be $9\cdot10^{-5}$. The residual UV reflection from the grating reaches $2\cdot10^{-5}$ when the diffraction angle is tuned for 148\,nm detection. Finally, a solar-blind PMT detection efficiency of about $5\cdot10^{-6}$ is measured with a calibrated light flux at 297\,nm. 
This corresponds to the total attenuation of $\geq\,10^{14}$ which is enough to suppress the UV background to the order of $10^3$\,counts per second. 

\begin{figure}[b]
\centering
\includegraphics[width=\linewidth]{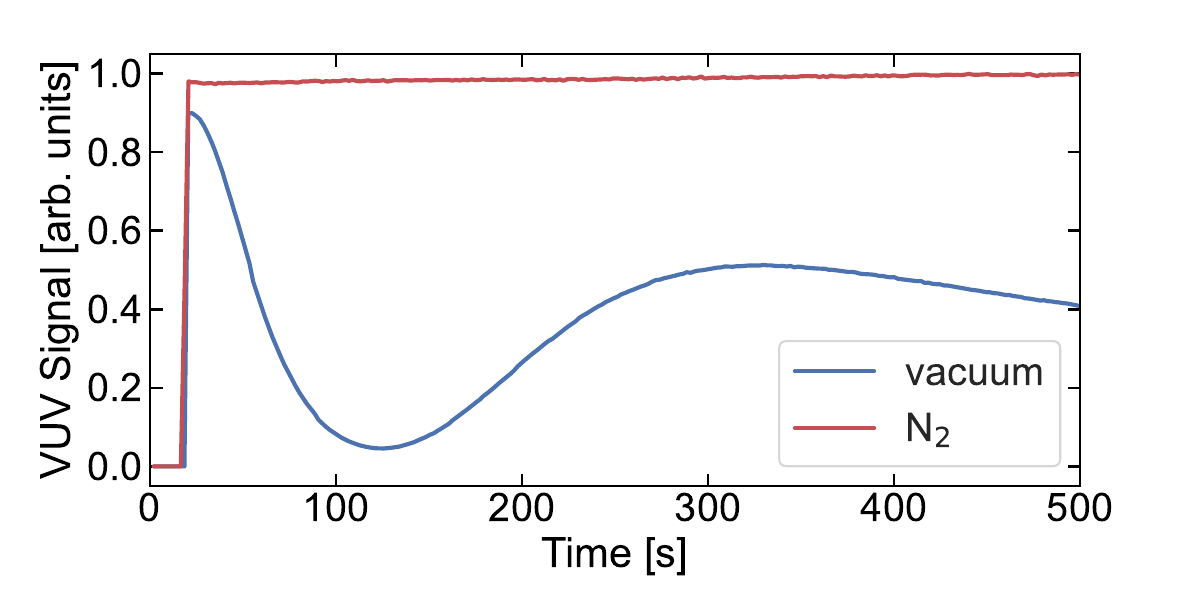}
\caption{ VUV signal measured with the SBO crystal in vacuum and under high-purity N$_2$ buffer gas. The laser radiation was initially blocked and then applied to the crystal. The crystal in vacuum shows strong local temperature variations that change the RQPM conditions and therefore the resulting VUV signal.}
\label{fig:n2_vs_vacuum}
\end{figure}

The overall VUV detection efficiency is calculated from the reflection coefficients of 83\,\% for p-polarization of the mirrors, a diffraction grating efficiency of 0.42 at 148\,nm, and a PMT quantum efficiency of 22\,\%   ($40.5\cdot10^{3}$ A/W anode radiant sensitivity) given by the suppliers.
The previously used MgF${_2}$ optical viewports and the collimation lens \cite{thielking2023vacuum}  have a transmission close to 90\,\% per element. 
The optical losses are mostly due to Fresnel reflection from uncoated surfaces. The overall VUV detection efficiency including the uncertainties of all the individual elements is 4(1)\,\%

The SHG power is derived from the PMT signal measured using two different detection methods.
We detect the signal by counting the VUV photons using a gated photon counter (counting rate up to 200\,MHz) and by PMT current measurements. 
The PMT signal discriminator for the gated counter is chosen in the  dark count plateau to avoid an overestimation of the VUV power.
However, when the photon counting rate exceeds 10 MHz we observe a saturation of the count rate determined by the pulse-pair resolution of the PMT. 
To overcome this effect 
for the high photon flux we measure the PMT current of the VUV signal by a picoammeter and take into account the anode radiant sensitivity as calibrated by the supplier for the specific PMT. We are using two different picoammeters  models and observe the same values with an uncertainty of $<$\,2\,\%.

The initial alignment of the spectrometer is performed with the fundamental radiation directed to the spectrometer by motorized VUV/UV separation mirrors.
After optimizing the UV signal on the PMT, the entire spectrum including the SHG signal can be recorded as shown in Fig.~\ref{fig:crystal_scan}. 
To discriminate against the scattering background and the occurrence of higher-order reflection of the fundamental radiation at the 148.4\,nm diffraction angle, a comparison measurement is also performed while the spectrometer is filled with air at atmospheric pressure. As expected, the peak at 148.4\,nm disappears due to the absorption of the VUV radiation in oxygen. 

\begin{figure}[h]
\centering
\includegraphics[width= 0.95\linewidth]{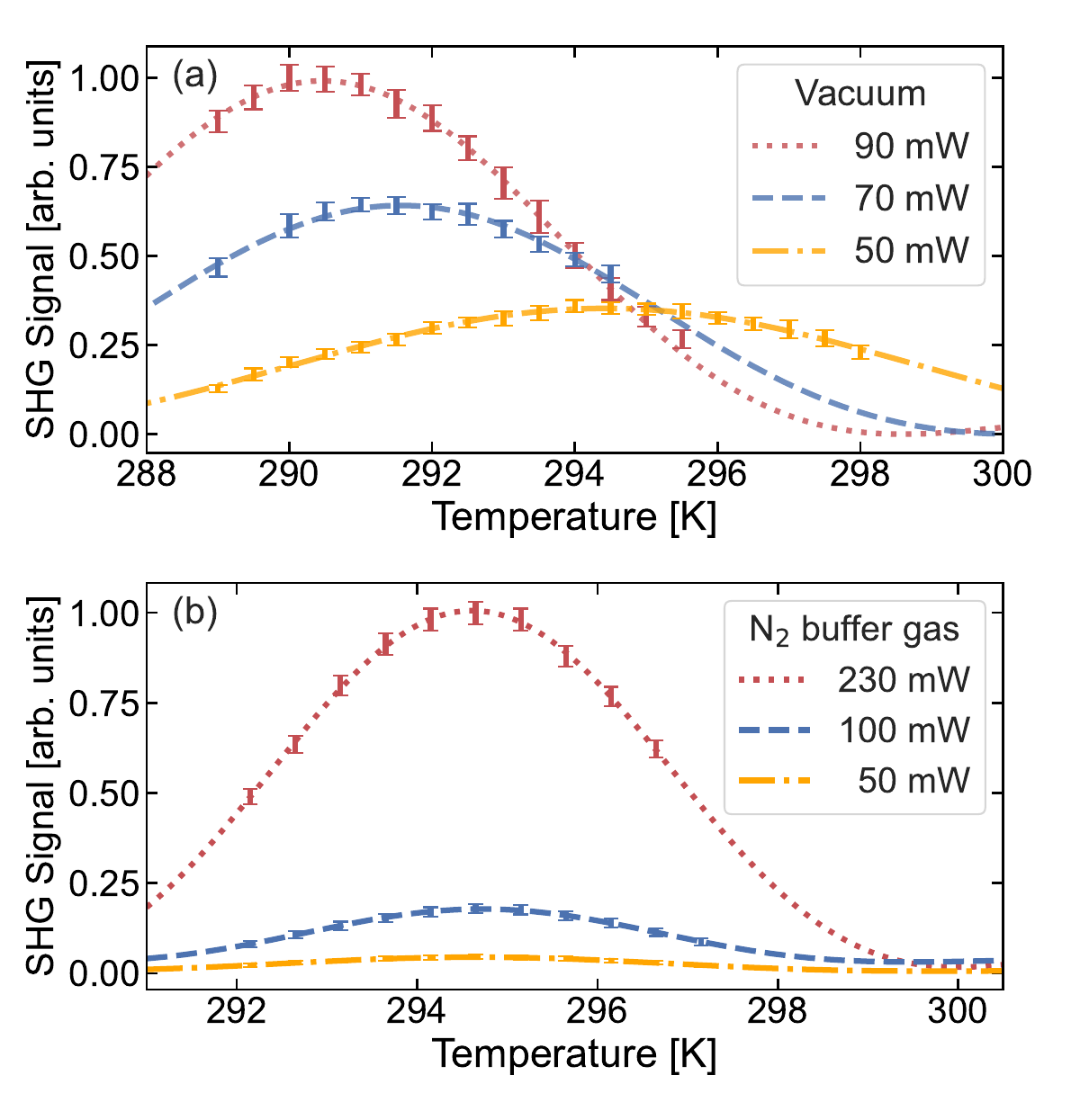}
\caption{\label{fig:temperature_dependence} SHG power as a function of crystal temperature for varying fundamental power under vacuum (a) and N$_2$ buffer gas environment (b).
The dashed lines are fits with sinc$^2$-functions. In the absence of nitrogen the optimal temperature for SHG decreases with increased fundamental power due to heating caused by absorption in the crystal.}
\end{figure}

The irregular domain structure of the SBO crystal is inhomogeneous across the sample. Therefore, we scan the crystal position relative to the laser beam to find the most efficient points. We found one area with the highest VUV output power and three areas with slightly lower VUV power.
The output power observed over the rest of the SBO is significantly lower. Each point requires fine adjustment of the incident UV beam to the crystal position to maximize the output SHG power. We also observe an angular dependence of the SHG power. Small rotations (changes of the effective nonlinear coefficient can be neglected) of the crystal around the \textit{b}- or \textit{c}-axes result in similar changes on the effective domain thickness and therefore SHG power. 
In our setup, a rotation around the crystal \textit{b}-axis is provided over a wider range. 
We have found that the optimal incidence angle is $\approx$\,9$^{\circ}$ in the \textit{a,c}-plane at room temperature. We assume this orientation better fits the RQPM conditions of the SBO crystal for the wavelength of 148.4\,nm. A similar behavior has been described before (e.g. \cite{aleksandrovsky2008}). 

For the determination of the optimal beam focusing, the output SHG power is measured as a function of the UV beam waist.   
We use an additional collimation lens placed after the crystal to avoid the influence of varying beam divergence on the detection.  
For the SBO sample used in this experiment we have found that the optimal UV beam waist radius in the crystal lies in the range of 20 - 25\,$\mu$m. 

\begin{figure}[h]
\centering
\includegraphics[width= 0.95\linewidth]{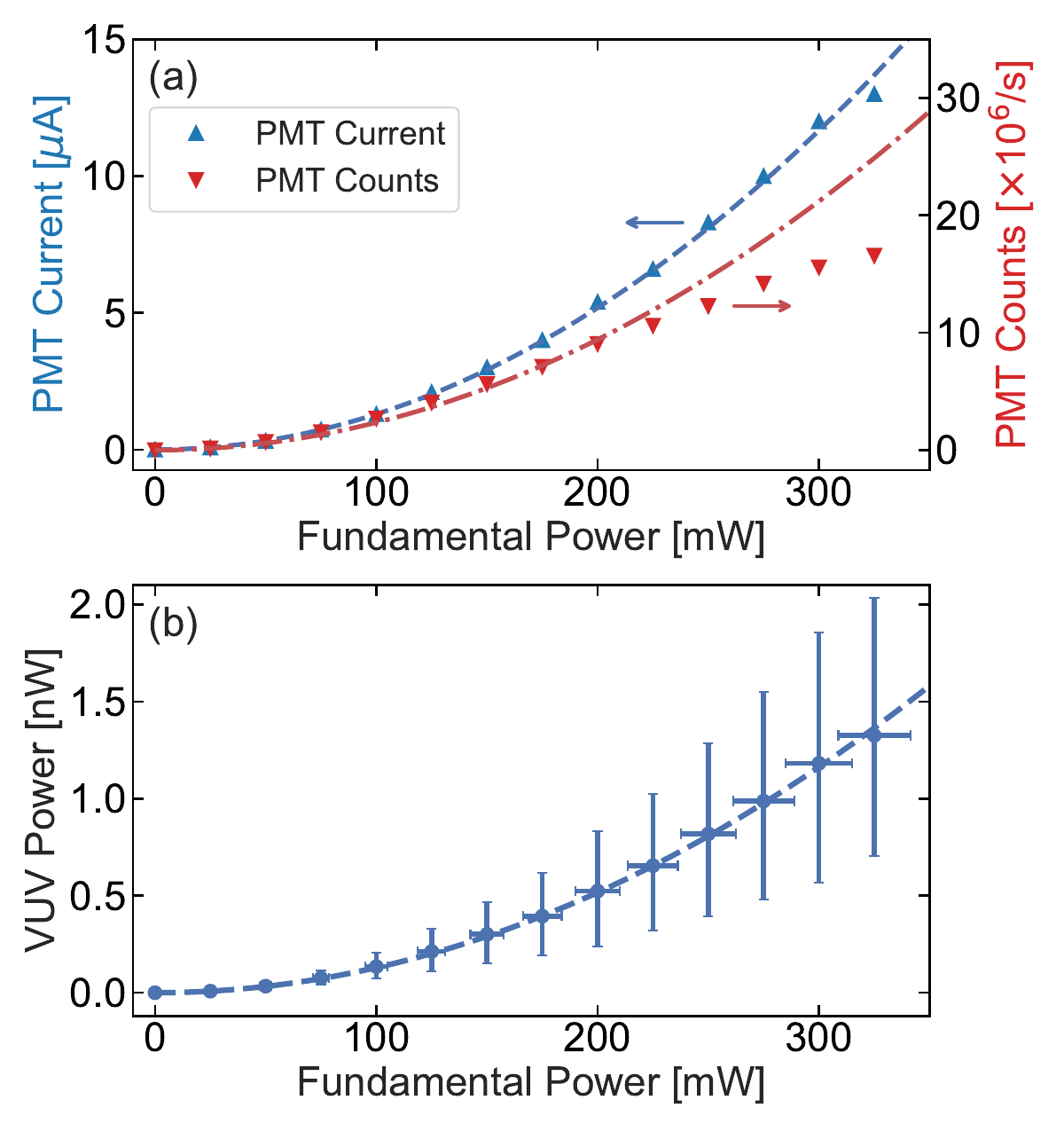}
\caption{\label{fig:power}
(a) Typical dependence of PMT VUV current and counts on the fundamental power with quadratic fits. For the photon counting dependence the fit ignores experimental points at UV powers $\ge$225\,mW where a saturation effect of the photon counting method is observed. 
(b) SHG power as function of the fundamental power.
}
\end{figure}

As mentioned above, the present SBO crystal exhibits some residual absorption at the fundamental wavelength. 
The local heating of the sample in vacuum causes thermal variations of the SHG power. We observe a fast reduction of the VUV power at high UV powers while keeping the setpoint of the temperature stabilization fixed. Filling the chamber with  high-purity N$_2$ gas, the output VUV power remains approximately constant over time  (see Fig.~\ref{fig:n2_vs_vacuum}). We assume that the buffer gas efficiently conducts heat away from the crystal surface and reduces the temperature gradients. This prevents changing of the RQPM conditions and the variation of the VUV power. The temperature dependence of the SHG power for different incident laser powers in vacuum and nitrogen buffer gas conditions is shown in Fig.~\ref{fig:temperature_dependence}.
Corrections of the crystal temperature are required for the compensation of the VUV power reduction in vacuum (see Fig.~\ref{fig:temperature_dependence}(a)). For an incident laser power of 250\,mW the optimal SBO operation temperature has to be reduced by $\approx\,10$\,K relative to the initial RQPM temperature at a laser power of $\leq\,50$\,mW.
In nitrogen buffer gas, this temperature correction is not required and the optimum operation temperature remains constant for a wide range of incident UV powers (see Fig.~\ref{fig:temperature_dependence}(b)).

The resulting SHG power in dependence on the fundamental incident power is shown in  Fig.~\ref{fig:power}. The process shows a quadratic behavior in the PMT current measurements and  the counting mode for UV powers below 225\,mW, as expected for the SHG process. Only for the counting method at high UV powers and a detected photon counting rate  $\geq$\,10$^7$\,s$^{-1}$ we observe a saturation of the detected counts as shown in  Fig.~\ref{fig:power}~(a). Therefore, the output power measured with photon counting is corrected for this effect. The measured output power is recalculated from the PMT signal using the detection efficiency. The uncertainties shown in the figure include both statistical and systematic contributions. The largest systematic contribution arises from the discrepancy between the two independent methods of measuring power and the uncertainty of the diffraction grating reflectivity (see Fig.~\ref{fig:power}~(b)). 
Scanning the laser frequency over $\approx$\,10\,GHz does not change the SHG power significantly, in agreement with previous calculations presented in \cite{aleksandrovsky2008}.

In conclusion, a CW VUV laser source  with the VUV power of 1.3\,$_{-0.6}^{+0.7}$\,nW at the excitation wavelength 148.4\,nm of the nuclear transition of $^{229}$Th has been demonstrated. Using an RQPM SBO crystal, we achieve a significantly  shorter CW wavelength  in comparison with the previously reported value of 191\,nm generated by phase-matched SHG in KBe${_2}$BO${_3}$F${_2}$ (KBBF) \cite{scholz20121}. 
The VUV power which is directed to a spectroscopy experiment (with Th-doped crystals or ions) is reduced by losses at a MgF$_2$ viewport and three mirrors.
Assuming a minimal VUV power after the crystal of $\approx\,0.7$\,nW and taking into account all optical losses, we expect to apply  $\geq\,0.3$\,nW  for investigations of the Th nuclear transition. 

The first laser excitation of the isomeric state in a $^{229}$Th-doped CaF$_2$ crystal was achieved using a VUV laser source based on four-wave frequency mixing in xenon \cite{thielking2023vacuum,tiedau2024laser} with the VUV power spectral density of $\approx$\,0.1 pW/Hz. 
With a CW VUV power of 0.3\,nW,  a linewidth in the kHz range would be required to reach a similar optical power density. 
For narrow linewidth high-precision spectroscopy a frequency stabilization system is now under development as employed in various optical clocks (see review \cite{ludlow2015optical}).

The main limitation of the present VUV laser source based on RQPM in strontium tetraborate is the unique domain structure of individual crystals and therefore, unique characteristics of the RQPM conversion efficiency.
The problem may be eventually resolved by manufacturing patterned SrB$_4$O$_7$ crystals \cite{perlov2024method}. 

\section*{Acknowledgements}
We would like to thank Thomas Leder, Martin Menzel, and Andreas Hoppmann for technical support.

This work has been funded by the European Research Council (ERC) under the European Union’s Horizon 2020 research and innovation programme (Grant Agreement No. 856415), the Deutsche Forschungsgemeinschaft (DFG) – SFB 1227 - Project-ID 274200144 (Project B04), and by the Max-Planck-RIKEN-PTB-Center for Time, Constants and Fundamental Symmetries.




\end{document}